\documentclass[aps,prl,twocolumn,superscriptaddress]{revtex4}

\usepackage[dvips]{graphicx}
\usepackage{color}
\usepackage{bm}
\usepackage{amssymb}
\renewcommand{\vec}[1]{\boldsymbol{#1}}
\usepackage[utf8]{inputenc}

\newcommand{\di}{\partial}

\begin{document}

\title{Anisotropic phase diagram and spin fluctuations of the hyperkagome magnet Gd$_3$Ga$_5$O$_{12}$ as revealed by sound velocity measurements}

\author{Alexandre Rousseau}
\author{Jean-Michel Parent}
\author{Jeffrey A. Quilliam}
\affiliation{Institut Quantique and Département de Physique, Université de Sherbrooke, Sherbrooke (Québec) J1K 2R1 Canada}

\date{\today}

\begin{abstract}
Sound velocity and attenuation measurements on the frustrated garnet material Gd$_3$Ga$_5$O$_{12}$ (GGG) are presented as a function of field and temperature, using two different acoustic modes and with two different magnetic field orientations: $[100]$ and $[110]$. We demonstrate that the phase diagram is highly anisotropic, with two distinct field-induced ordered phases for $H||$[110] and only one for   $H||$[100]. Extensive lattice softening is found to occur at low fields, which can be associated with spin fluctuations. However, deep within the spin liquid phase a low-temperature stiffening of the lattice and reduced attenuation provides evidence for a spin gap which may be related to short-range antiferromagnetic correlations over minimal 10-spin loops.

 \end{abstract}


\maketitle

An understanding of the low-temperature properties of Gd$_3$Ga$_5$O$_{12}$ (GGG) is one of the longest outstanding problems in the field of geometrically frustrated magnetism. This material is thought to consist of large, classical spins ($J=S=7/2$) interacting antiferromagnetically on two interpenetrating hyperkagome lattices. Such a situation could be expected, on a classical level, to maintain a high degree of frustration and exhibit some kind of classical spin liquid at low temperatures. However, the dipolar interaction between the large Gd moments is also very relevant and would normally be expected to relieve the frustration in this material. Hence the fact that GGG does not develop long range order (LRO) in zero-field is quite surprising. 

Instead, it has been found to undergo unconventional spin freezing, into a state that incorporates glassiness~\cite{Schiffer1995,Ghosh2008}, extended short-range order~\cite{Petrenko1998,Yavorskii2006,Yavorskii2007JPCM} and persistent spin fluctuations~\cite{Marshall2002,Dunsiger2000}. It is tempting to simply attribute such a state to quenched crystalline disorder since GGG is known to have a $\sim$1\% off-stoichiometry~\cite{Daudin1982} and other Gd-garnets show LRO~\cite{QuilliamGarnets,Applegate2007}. However, it has also recently been demonstrated, albeit on a different lattice, that such an inhomogeneous ``spin slush'' can in principle occur in a structurally clean system~\cite{Rau2016}. With the application of a small magnetic field, the frozen magnetism in GGG is melted and the system enters an apparent spin liquid regime~\cite{Schiffer1994,Tsui1999}. The nature and origin of these exotic phases remain mysterious despite more than 20 years of intermittent research on the problem. The most recent development in the field has been the observation of antiferromagnetic correlations on 10-spin rings, described by a nematic order-parameter, or director~\cite{Paddison2015}. 

At higher magnetic fields, a bubble of field-induced antiferromagnetism is observed~\cite{Schiffer1994}. Later work~\cite{Petrenko2009,Deen2015} showed that this bubble in fact consisted of two distinct phases of LRO with different symmetries, although the precise magnetic structure of these phases has yet to be refined. Furthermore, there remains some inconsistency regarding the phase boundaries and their dependence on field direction and sample geometry~\cite{Petrenko2009,Deen2015,Schiffer1995}. Concrete progress in understanding this material is unlikely to be achieved until a definitive $H$-$T$ phase diagram is established, thus we have embarked on a comprehensive study of the phase diagram of GGG using two different field orientations. We have applied the technique of sound velocity measurements which provides a highly sensitive probe of magnetic phase transitions. Moreover, the magneto-elastic coupling permits us to study the evolution of spin fluctuations, particularly in the perplexing spin liquid phase~\footnote{Here, we use the term ``spin liquid'' to describe the mysterious lack order or freezing in this region of the phase diagram rather than to invoke a true quantum spin liquid state.}.

\begin{figure}
\begin{center}
\includegraphics[width=3.35in,keepaspectratio=true]{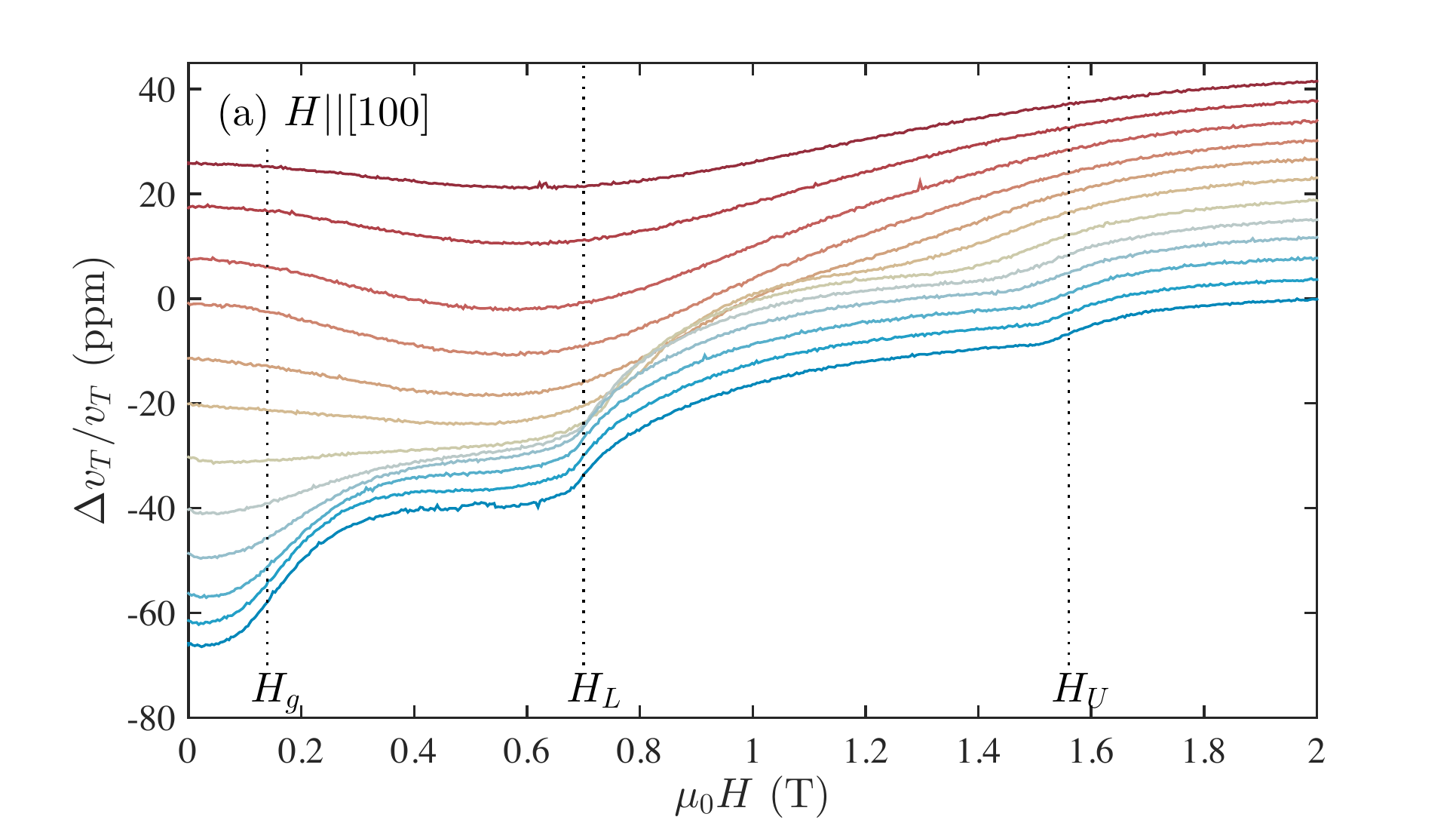}
\includegraphics[width=3.35in,keepaspectratio=true]{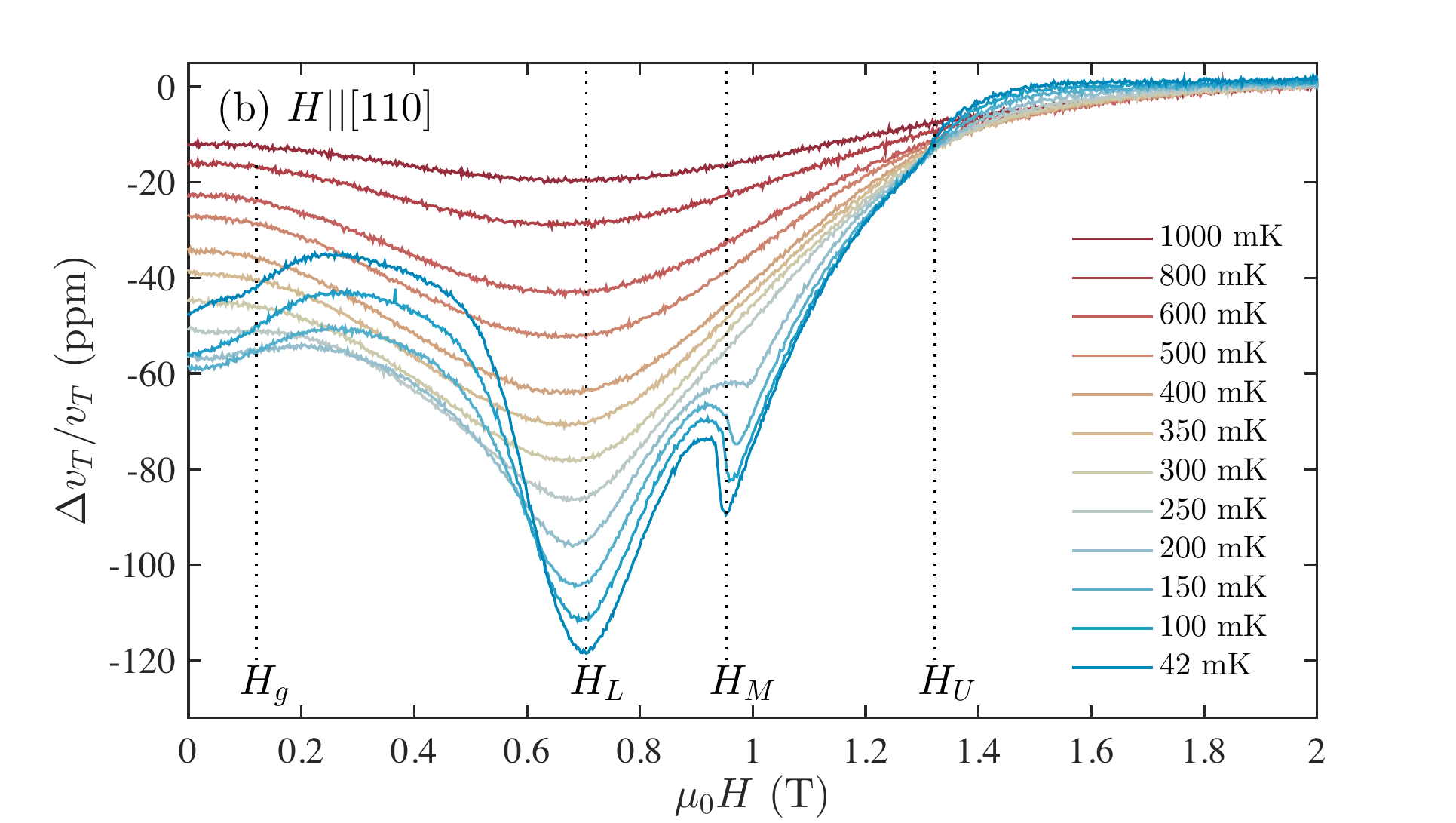}
\caption{Field scans of the relative change in transverse sound velocity, $\Delta v_T/v$, for various temperatures. The field is oriented along the $[110]$ direction in (a) and along the $[100]$ direction in (b). In (b) the curves are staggered for ease of view. The same legend applies to panels (a) and (b).  \label{Hscans}}
\end{center}
\end{figure}

A large GGG single crystal substrate was purchased from \emph{MolTech Berlin} and was cut to dimensions 8.7 mm $\times$  4.9 mm $\times$ 1.1 mm along [110], [1$\bar{1}$0] and [001] directions respectively. LiNbO$_3$ piezoelectric transducers were glued to parallel surfaces so as to propagate sound waves along $\vec{k} \propto$ [1$\bar{1}$0]. A transverse acoustic mode was studied with $\vec{u} ||$[110], which has a sound velocity $v_T = \sqrt{(C_{11} - C_{12})/2\rho} \simeq 3680$ m/s at ambient temperature. The longitudinal mode with the same direction of propagation was found to be weakly coupled to the magnetic degrees of freedom. The sample was affixed to the mixing chamber of a dilution refrigerator and cooled to as low as 40 mK. Two magnetic field orientations were studied, along [110] (perpendicular to $\vec{k}$) and along [100] (at a 45$^\circ$ angle to $\vec{k}$). Relative changes in sound velocity, and therefore relative changes in elastic constant, were measured using an ultrasonic interferometer. The frequency was varied so as to maintain a constant phase between transmitted pulses and the echoes, allowing us to measure $\Delta v_T / v_T = \Delta f / f $.

In rare-earth materials, variations in sound velocity are typically coupled to the magnetic degrees of freedom through two main mechanisms. (1) A lattice strain can modify the crystal field which couples to the total angular momentum via Stevens' operators~\cite{Luthi}. For Gd$^{3+}$ in GGG, this may be a secondary effect since to a first approximation, $L=0$, $J = S = 7/2$ and the crystal field has a smaller effect than in many rare-earth systems. (2) In the case of  exchange-striction, the interactions between adjacent magnetic moments are dependent on the distance between them, leading to a coupling between lattice strain and the magnetic moments~\cite{Luthi}. In either case, to first order, relative changes in sound velocity are related to the square of an order parameter $(M^\dag)^2$ or the uniform magnetization $M^2$. The magnitude and sign of the coupling constants depend on the symmetry of the magnetic structure and the acoustic mode. Spin fluctuations may also lead to an appreciable softening of acoustic modes~\cite{Quirion2015}.

{\bf Phase boundaries and anisotropy} – Beginning with the field oriented along the [100] direction at the lowest temperatures, $\Delta v_T/ v_T$ shows three distinct anomalies (identified by a maximum in slope, $\partial v_T/\partial H$), as shown in Fig.~\ref{Hscans}(a). Given previous work on this system~\cite{Schiffer1994} and the fact that this change in slope is rather gradual, we identify the first anomaly at $\mu_0H_g \simeq 0.14$ T as the point at which the unconventional spin glass (SG) freezing is suppressed, giving way to the spin liquid phase. Two other sharp changes of slope are observed and the critical fields $H_L = 0.70$ T and $H_U = 1.56$ T are identified by a maximum in $\partial v_T/\partial H$ as indicated in Fig.~\ref{Hscans}(a). These anomalies appear to be the lower ($H_L$) and upper ($H_U$) phase boundaries of field-induced antiferromagnetic order and are quite consistent with the results of Schiffer \emph{et al.}~\cite{Schiffer1994}, which were taken with the same magnetic field orientation. Compared to the magnetization~\cite{Schiffer1994}, the sound velocity provides much sharper changes in slope at the transitions, which indicates a coupling between the strain and an antiferromagnetic order parameter, $M^\dag$, for $H_L < H < H_U$. Since the sound velocity continues to gradually increase and then saturates above $H_U$, we expect that there is an additional coupling to the square of the uniform magnetization, $M^2$.  Based on the slope above $H_U$, in the field-polarized paramagnetic regime, roughly 30 ppm of variation over the whole field range can be attributed to the uniform magnetization. Other variations in velocity are likely a result of spin fluctuations.

 \begin{figure*}
\begin{center}
\includegraphics[width=2.3in,keepaspectratio=true]{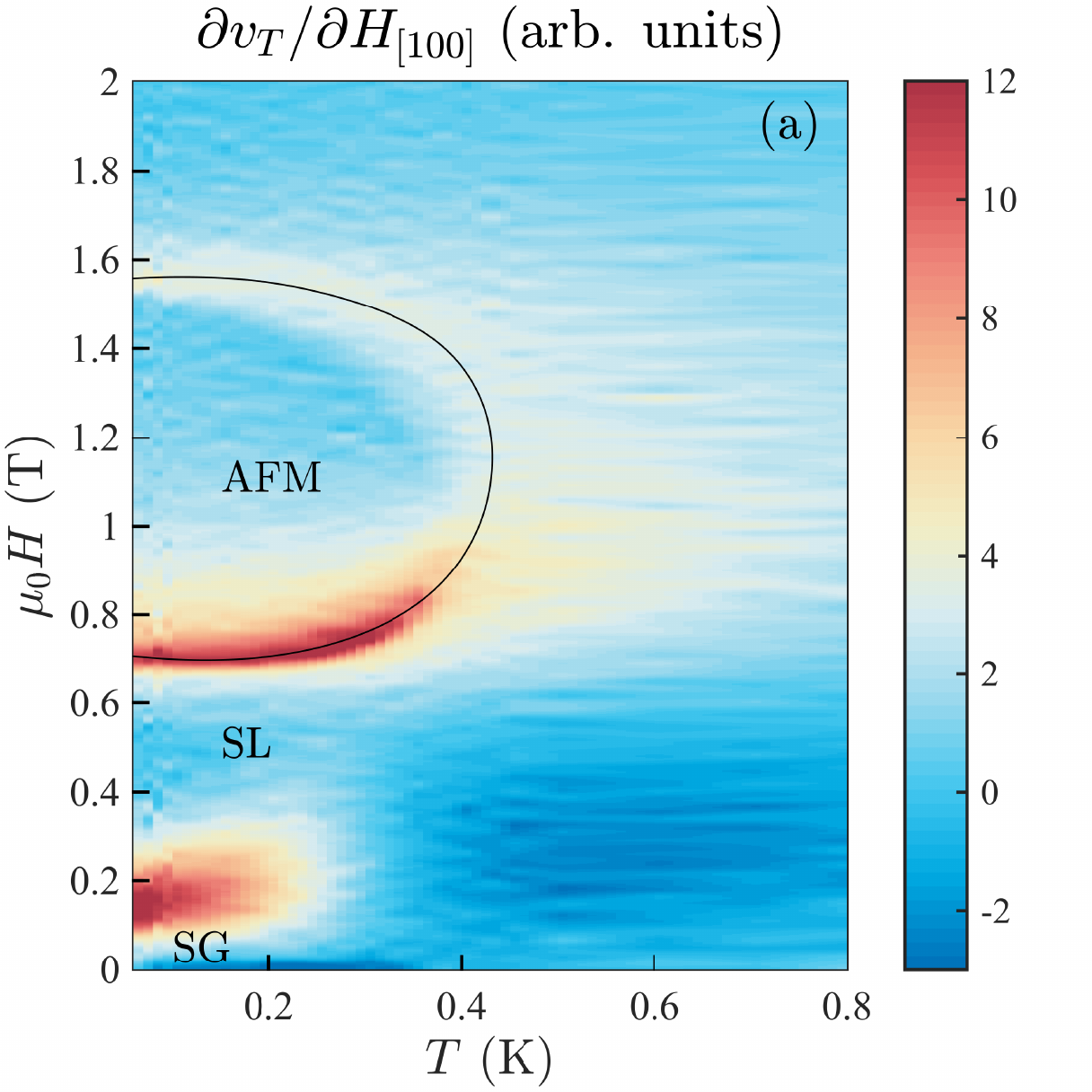} 
\includegraphics[width=2.3in,keepaspectratio=true]{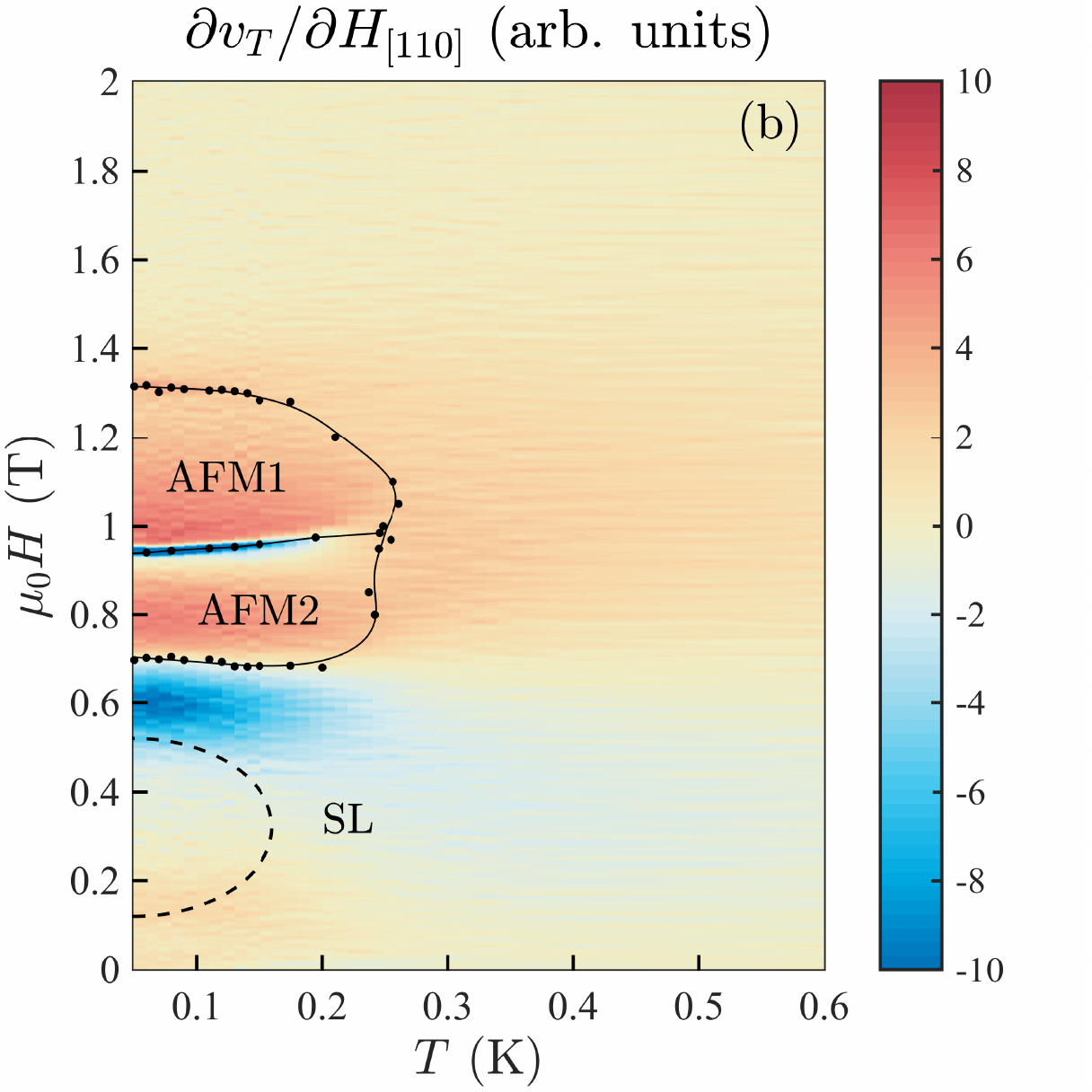} 
\includegraphics[width=2.3in,keepaspectratio=true]{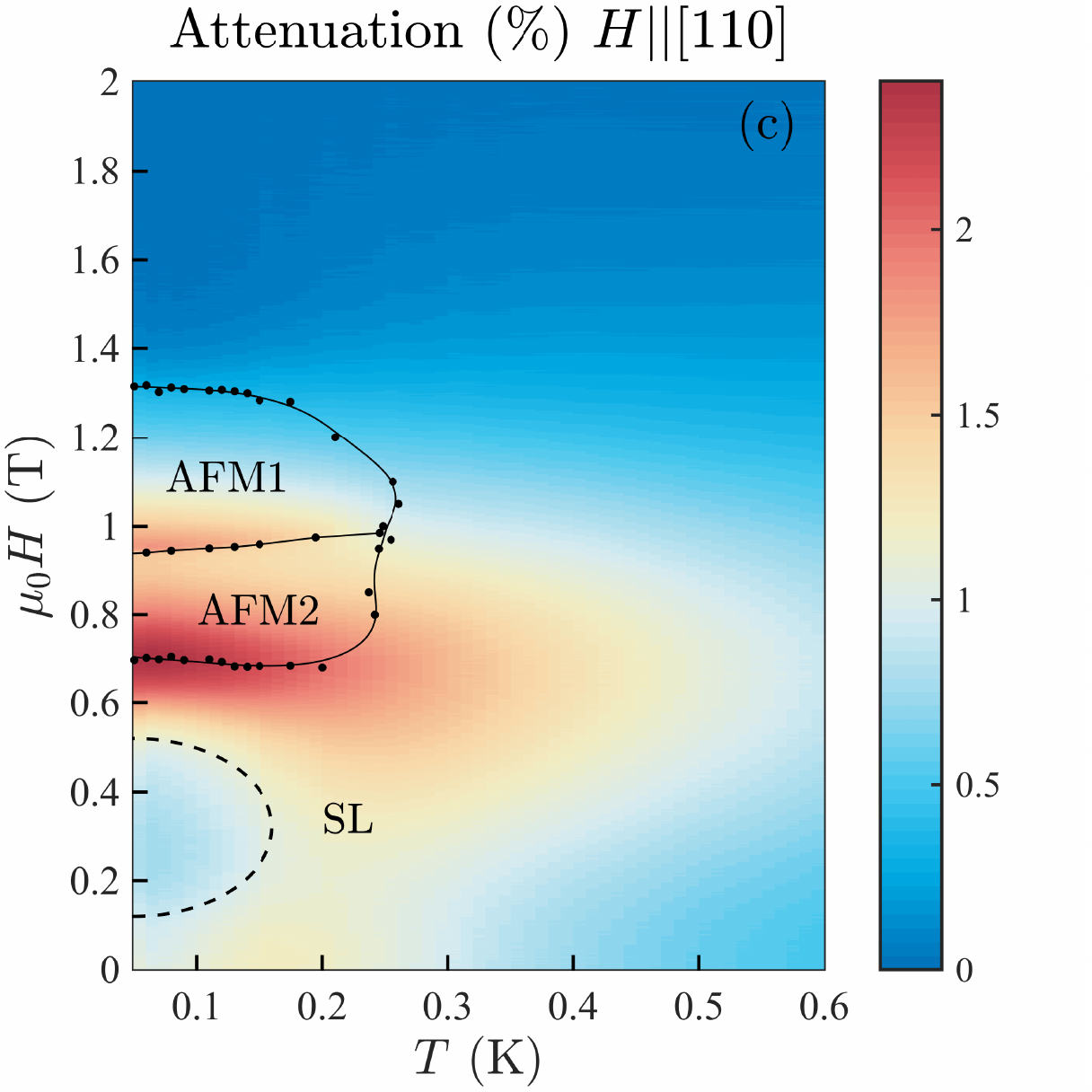}
\caption{(a) A surface plot of $\di v_T / \di H$ for $H||[100]$ with the field-induced antiferromagnetic phase indicated with a black line. The boundary between spin glass (SG) and spin liquid (SL) phases is also evident as a maximum in $\di v_T / \di H$. (b) A surface plot of $\di v_T / \di H$ for $H||[110]$. The phase boundaries, identified by black points, are elucidated from both field and temperature scans, as described in the text, and show two distinct antiferromagnetic phases (AFM1 and AFM2). (c) Relative sound attenuation (percent decrease in signal relative to the value at $H=2$ T, $T=50$ mK). A pronounced region of decreased attenuation and lattice stiffening within the spin liquid phase, which likely indicates gapped excitations, is highlighted with a dashed line. \label{Surfaces}}
\end{center}
\end{figure*}

With the field applied along the [110] direction, the $\Delta v_T/v_T$ curves change dramatically, as shown in Fig.~\ref{Hscans}(b). For this field orientation, four different anomalies are observed. A subtle change in slope, likely signalling a departure from spin glassiness is seen at $H_g\simeq 0.12$ T, roughly the same field as for the [100] orientation. The onset of antiferromagnetic order at $H_L = 0.705$ T coincides with a very deep minimum in the sound velocity. The upper range of antiferromagnetic order is found at $H_U = 1.32$ T at a subtle change of slope, similar to that observed for [100]. What is most different about the [110] orientation is the appearance of an additional and very sharp transition at $H_M = 0.95$ T.  Hence we demonstrate clear evidence of two distinct antiferromagnetic phases: phase AFM1 between $H_L$ and $H_M$ and phase AFM2 from $H_M$ to $H_U$. These two distinct antiferromagnetic phases have previously been observed by Petrenko \emph{et al.}~\cite{Petrenko2009} in neutron diffraction measurements and by Deen \emph{et al.}~\cite{Deen2015} by neutron scattering and magnetic susceptibility measurements. However, this work on two different orientations of the same sample, using the same technique, makes clear that two distinct AFM phases are only observed for $H||$[110] and not for $H||$[100].

For the [100] field-orientation the evolution of phase boundaries as a function of temperature is fairly easy to follow. As shown in Fig.~\ref{Surfaces}(a), local maxima of a surface plot of $\partial v_T / \partial H$ vs. $H$ and $T$, clearly show the limits of the antiferromagnetic ``bubble''. These phase boundaries agree fairly well with Schiffer \emph{et al.}~\cite{Schiffer1994}, although we do not observe the same reentrance near $H_L$ as was highlighted by Tsui \emph{et al.}~\cite{Tsui1999}. Our phase boundaries appear to occur at slightly higher magnetic field than those of Schiffer \emph{et al.}~\cite{Schiffer1994} (0.68 and 1.507 T) but this is likely just a result of demagnetization (since the demagnetizing field reduces the internal field, higher applied fields are needed to induce a transition).

\begin{figure}
\begin{center}
\includegraphics[width=3.35in,keepaspectratio=true]{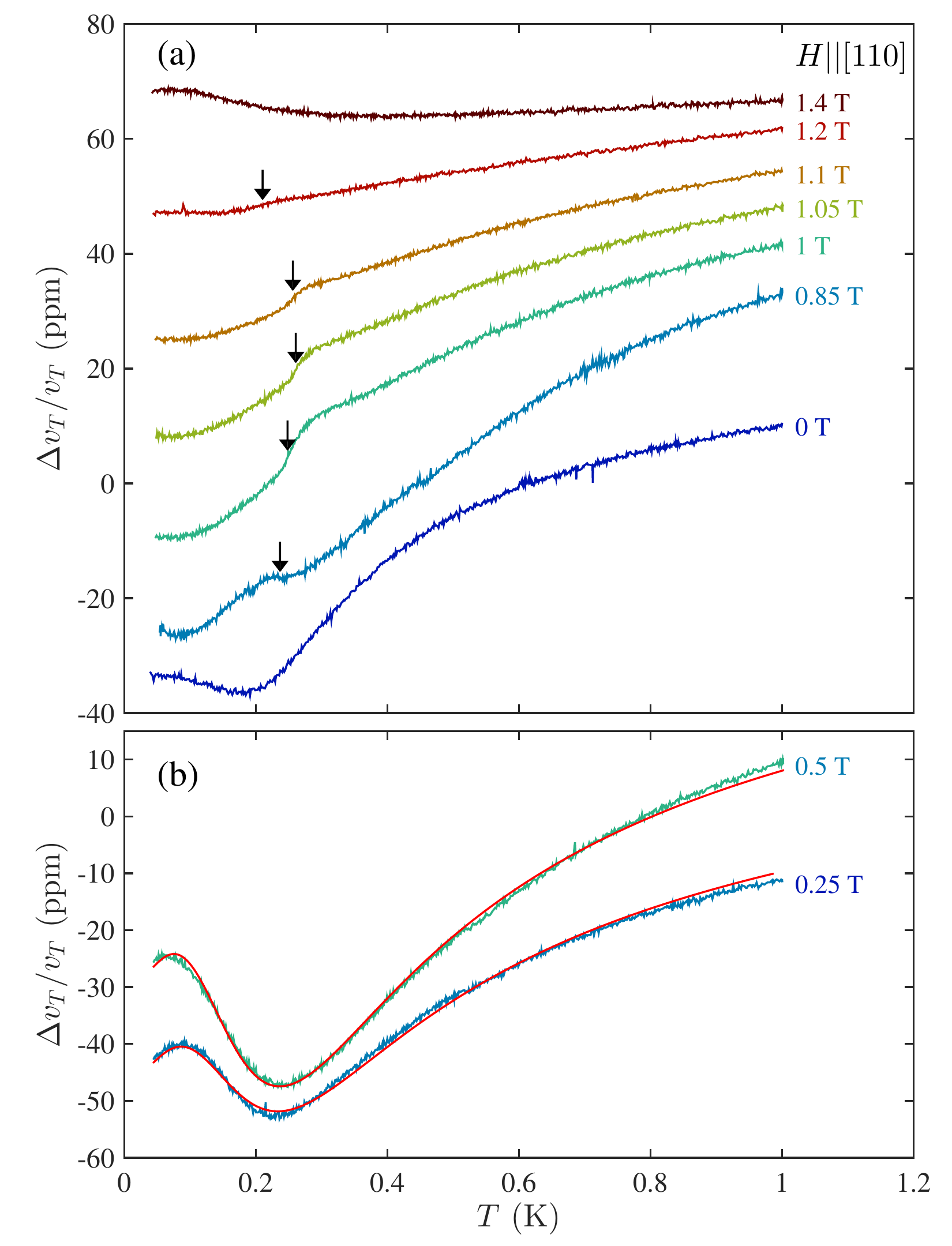}
\caption{Temperature scans of the relative change in transverse velocity at various magnetic fields, (a) in zero-field and in the region of field induced order and (b) in the spin liquid phase. Ordering anomalies are indicated with black arrows. Red curves in (b) are fits of the data at 0.25 and 0.5 T, as described in the text. \label{Tscans}}
\end{center}
\end{figure}

For the [110] field-orientation, the transitions in field are easily observed, but transitions in temperature are much less apparent, as shown in Fig.~\ref{Surfaces}(b). With more detailed temperature scans, small anomalies are detected, as indicated with arrows in Fig.~\ref{Tscans}(a), allowing us to draw the $H$-$T$ phase lines in Fig.~\ref{Surfaces}(b). This phase diagram is fairly consistent with that of Deen \emph{et al.}~\cite{Deen2015} with just a $\sim 10$\% difference in field values. On the other hand, the neutron scattering data of Ref.~\cite{Petrenko2009} indicate phase boundaries at much higher magnetic fields: $H_L \simeq 0.9$ T, $H_M \simeq 1.25$ T and $H_U \simeq 1.7$ T.  Again, these discrepancies could likely be explained by the demagnetization factor. Additionally, Deen \emph{et al.}~\cite{Deen2015} have identified various anomalies delimiting short range order higher in temperature than the LRO antiferromagnetic phases. We do not observe corresponding anomalies in sound velocity.

{\bf Spin liquid phase} – As expected from previous studies, the spin liquid phase does not exhibit any sharp anomalies, implying that it is indeed an order-free dynamic phase. However, this phase is far from featureless. As shown in the curve at 0.5 T in Fig.~\ref{Tscans}(b), for example, there is a significant softening of the lattice as the temperature is reduced until around 0.25 K, at which point the velocity passes through a broad minimum and begins to increase again. A softening of the lattice due to spin fluctuations just above a phase transition is a common observation~\cite{Quirion2015}. However, the observed broad minimum and the absence of a phase transition is more unusual and is generally an indication of an energy gap in the spin-excitation spectrum~\cite{Zherlitsyn2000,Wolf2001}. To qualitatively understand this behavior, it is sufficient to suppose that there is a strain dependence to the energy gap $\Delta(\epsilon)$ which gives rise to a softening of the lattice in the vicinity of $T\simeq \Delta/k_B$. Below the gap temperature, as excitations become much rarer, the original, high-temperature elastic constants are recovered.

Inspired by very similar temperature variations of the sound velocity that have been observed in the frustrated spinel compounds MgCr$_2$O$_4$ and ZnCr$_2$O$_4$~\cite{Watanabe2012}, we will attempt to understand our results through the coupling of weakly interacting spin clusters to the lattice strain. In such $B$-site spinels the magnetic ions occupy a pyrochlore lattice, a highly frustrated 3d network of corner-sharing tetrahedra. After the tetrahedra, the next geometric forms of importance are 6-spin loops. While ZnCr$_2$O$_4$ exhibits a magnetic phase transition at $T_N\simeq 13$ K, there is a broad, correlated ``liquid'' regime above that point~\cite{Watanabe2012}. Rather than a collection of paramagnetic spins on a pyrochlore lattice, the diffuse neutron scattering profile~\cite{Lee2002ZnCr2O4} in this correlated regime is found to be representative of weakly interacting ``directors'', which are defined by the staggered magnetization around a loop as
\begin{equation} \vec{L}(\vec{r}) = \frac{1}{N} \sum_{n=1}^N (-1)^n \vec{S}_n(\vec{r}) \end{equation}
where $N=6$ in the case of the pyrochlore lattice. The correlated regime can therefore be viewed as an assembly of largely uncorrelated, emergent molecular magnets or spin clusters. Similar closed loops are thought to play an important role in many frustrated magnetic systems where uniformly rotating or flipping spins on particular closed paths often does not change the energy of the system, at least at the classical level. Note that neutron scattering measurements on the pyrochlore spin ice material Dy$_2$Ti$_2$O$_7$ also showed results that are nearly consistent with uncorrelated hexagonal loops, although a much better agreement with the data was obtained with extensive Monte Carlo simulations that incorporate further-neighbor exchange interactions into the usual dipolar spin ice Hamiltonian~\cite{Yavorskii2008}. 

A very similar situation to that of ZnCr$_2$O$_4$ has been proposed for GGG by Paddison \emph{et al.}~\cite{Paddison2015}. Having analyzed diffuse neutron diffraction data in zero field, just above the spin glass transition, they show that the spin correlations in GGG can also be described by directors, this time defined by staggered magnetization around 10-spin loops (the next shortest closed path after the triangular plaquettes). The orientation of the directors $\vec{L}$ at low $T$ is found to be predominantly along the local $z$-axis, though the sign of $L_z$ remains random. It has also been noted in Ref.~\cite{Ambrumenil2015} that at high-field in a nearest-neighbor model, the excitations out of the field-polarized state are localized on the 10-spin loops and result in flat bands.

Just as observed here for GGG, sound velocity measurements of ZnCr$_2$O$_4$ and MgCr$_2$O$_4$, in the correlated paramagnetic regime, show a softening and broad minimum~\cite{Watanabe2012}. Watanabe \emph{et al.} have associated the lattice softening, and hardening at lower temperatures, with the strain dependence of an energy gap required to break these 6-spin correlations and have fit their data following the analysis of Wolf \emph{et al.}~\cite{Wolf2001}, who considered the $T$-dependence of sound-velocity in the gapped dimer-system SrCu$_2$(BO$_3$)$_2$. Here we follow a similar scheme to explain the related phenomenology in GGG.

A minimal model that considers clusters with a ground state manifold, including $|L_z = -S \rangle$ and $|L_z = +S\rangle$, and an excited manifold at energy $\Delta$ with relative multiplicity $\alpha$, gives the partition function $Z = 1 + \alpha e^{-\beta \Delta}$. It is also supposed that a lattice strain $\epsilon$ alters $\Delta$ via exchange-striction. To leading order, we can expect a linear coupling $G = \partial \Delta/\partial\epsilon > 0$ for the acoustic modes studied here. In a mean-field approach, the $q=0$ elastic constant is given by $C(T) = C_0 - NG^2 \chi_S/(1-K\chi_S)$ where $\chi_S$ is a local strain susceptibility for a single cluster (or dimer), $K$ is a coupling strength between clusters and $N$ is the density of spin clusters~\cite{Wolf2001}. The local strain susceptibility is simply related to the free energy $F$ of a single cluster by $\chi_S = -\frac{1}{G^2}\frac{\partial^2 F}{\partial  \epsilon^2}$. This leads to the following expression for relative changes in sound velocity:
\begin{equation}
\frac{\Delta v_\mathrm{gap}}{v_0} = - \frac{(NG^2\alpha/2C_0) e^{-\Delta/T}}{ T(1 + \alpha e^{-\Delta/T})^2 - K\alpha e^{-\Delta/T}} 
\label{gap}
\end{equation}
Although this function produces an appropriately broad minimum in sound velocity, our data show that the high-temperature elastic constant is not entirely recovered at low temperatures, suggesting that there is an additional component that is not fully ``gapped out'' by 50 mK. Hence we have also included a Curie-Weiss background contribution, $\Delta v_\mathrm{CW}/v_0 = A_\mathrm{CW}/(T - \Theta)$, that is simply the $T > \Delta$ limit of Eq.~\ref{gap} and therefore represents gapless (or very weakly gapped) spin fluctuations. Our final fitting function is given by the sum of these two contributions, that is $\Delta v_T = \Delta v_\mathrm{gap} + \Delta v_\mathrm{CW}$. This additional background softening $\Delta v_\mathrm{CW}$ could come from various sources. It may be that not all 10-spin loops are gapped as the applied field makes for several inequivalent loops with different crystallographic orientations. Alternatively, fluctuations between the degenerate (or nearly so)  $|L_z = +S\rangle$ and $|L_z = -S\rangle$  ground states of each cluster may explain a lattice softening that continues to very low temperatures. Presumably, the absence of this background softening would imply a system with a completely frozen ground state and would be inconsistent with a spin liquid phase. 

Fits of this phenomenological model, using $\alpha=2$, are shown in Fig.~\ref{Tscans}(b) for select values of magnetic field. A value of $\alpha=2$ may be rationalized by supposing that the minimal excitations of the 10-spin loops ressemble $k=0$ magnons of a chain of spins with cyclic boundary conditions, giving rise to two distinct modes. We have used fitting parameters $K$, $\Theta$ and $C_0$ that are independent of field (since these parameters are non-magnetic in origin) while $\Delta$, $NG^2$ and $A_\mathrm{CW}$ are allowed to vary freely. The value of gap that we extract is $\Delta = 0.52(1)$ K with no statistically significant variation between 0.25 T and 0.5 T. The strain coupling between gapped clusters is found to be ferrodistortive, with $K \simeq 0.35$ K, whereas $\Theta \simeq -0.57$~K indicates an antiferrodistortive coupling for the gapless contribution. The same analysis can be applied to the [100] field-orientation, though there the gap is found to have more variability as a function of magnetic field.

It is also beneficial to study the effect of field and temperature on relative changes in sound attenuation, as shown in Fig.~\ref{Surfaces}(c). Notably, the softening of the lattice at low temperature and low field is found to be correlated  with an increase in sound attenuation. Increased sound absorption due to spin fluctuations is a well-known phenomenon~\cite{Laramore1969,Suzuki1980}, especially in the vicinity of a phase transition. Interestingly we also find a region of \emph{decreased} attenuation at low temperature within the spin liquid phase, highlighted with a dashed line in Figs.~\ref{Surfaces}(b) and (c). This decrease in attenuation provides another indication of the gap, below which there are fewer spin excitations present to dissipate the sound waves. 

In conclusion, these results have led us to two principal conclusions regarding the various low-temperature phases of GGG. First, we have demonstrated unambiguously that the phase diagram is highly anisotropic, with two field-induced antiferromagnetic phases for $H||[110]$ and only one antiferromagnetic phase for $H||[100]$. This realization may help to constrain theoretical and experimental work aiming to understand the physics of GGG, as concentrating on well-defined ordered phases may be more straightforward than directly tackling the enigmatic spin liquid phase at lower field. As a first approximation the Gd ions have a half-filled shell giving $L=0$ and isotropic spins, $J=S=7/2$. However, a single-ion anisotropy of 40 mK has previously been inferred~\cite{Deen2015} and is not insignificant relative to the nearest-neighbour exchange interaction $\mathcal{J} = 107$ mK)~\cite{Kinney1979}. Moreover, one should not ignore the fundamental importance of the long-range dipolar interaction ($D=45$ mK at the nearest-neighbor level~\cite{QuilliamGarnets}) which also generates significant anisotropies~\cite{Quilliam2007GSO,DelMaestro2007} and may be responsible for the anisotropy of field-induced phases. 

Secondly, our measurements provide evidence of a spin gap in the excitation spectrum of the spin liquid phase. Additional softening of the lattice at the lowest temperatures implies that there is nonetheless a rich spectrum of spin fluctuations below the gap. It remains to be proven that this gap is truly associated with the emergent 10-spin directors invoked by Paddison \emph{et al.}~\cite{Paddison2015} and whether it somehow plays a role in stabilizing the spin liquid ground state. The work of Paddison \emph{et al.}~\cite{Paddison2015} was carried out at fairly high temperatures (175 mK) and in zero-field, perilously close to spin glass freezing~\cite{Schiffer1995}, hence it would be very valuable to repeat such experiments and analysis in the middle of the spin liquid regime where we have noted such a clear decrease in lattice softening and sound attenuation.

\begin{acknowledgements}
We are grateful to M. Gingras and G. Quirion for a critical reading of the manuscript and acknowledge technical support from M. Castonguay and other members of the technical staff at the Institut Quantique. This work was supported by NSERC and the FRQNT. 
\end{acknowledgements}

\end{document}